\documentclass{PoS}

\title{Bridge Programs as an approach to improving diversity in physics}

\ShortTitle{Bridge Program in Physics}

\author{\speaker{Brian Beckford}
        University of Michigan\\
        E-mail: \email{bobeck@umich.edu}}

\abstract{In most physical sciences, students from underrepresented minority (URM) groups constitute a small percentage of earned degrees at the undergraduate and graduate levels. Bridge programs can serve as an initiative to increase the number of URM students that gain access to graduate school and earn advanced degrees in physics. This talk discussed levels of representation in physical sciences as well as some results and best practices of current bridge programs in physics. The APS Bridge Program has enabled over 100 students to be placed into Bridge or graduate programs in physics, while retaining 88\% of those placed.}

\FullConference{38th International Conference on High Energy Physics\\
		3-10 August 2016\\
		Chicago, USA}

\begin{document}
\section{Introduction}
Physics has long been thought of as the leader of physical sciences and at the forefront of scientific discovery. However, when we consider the representation of groups in our society, we do not see the same, or even close representations in the field of physics. In fact, physics is one of the least diverse of all physical sciences with respect to gender, underrepresented minorities, and the LGBTQ+ community. The situation has not improved over time, in fact the number of degrees earned by African Americans has decreased in the last decade. Leaders at physics organizations and universities have recognized the demand for improved diversity. This has prompted official diversity statements from the American Physical Society (APS) and the American Association of Physics Teachers (AAPT). More recently was the creation of the APS forum on Diversity and Inclusion slated to begin in 2017. In response to the need for addressing the level of representation on URM in physics, the APS Department of Education and Diversity pioneered the APS Bridge Program, a national effort to increase the number of physics PhDs awarded to underrepresented minority (URM) students, defined by the project as African Americans, Hispanic Americans and Native Americans. There are numerous approaches targeted to improving diversity in physics but in this report, we present the best practices of Bridge Programs in physics and some initial outcomes of institutions that have adopted these programs. 

\section{Motivation}
  \begin{figure}[htb]
	\begin{center}
		\includegraphics*[width=.49\columnwidth]{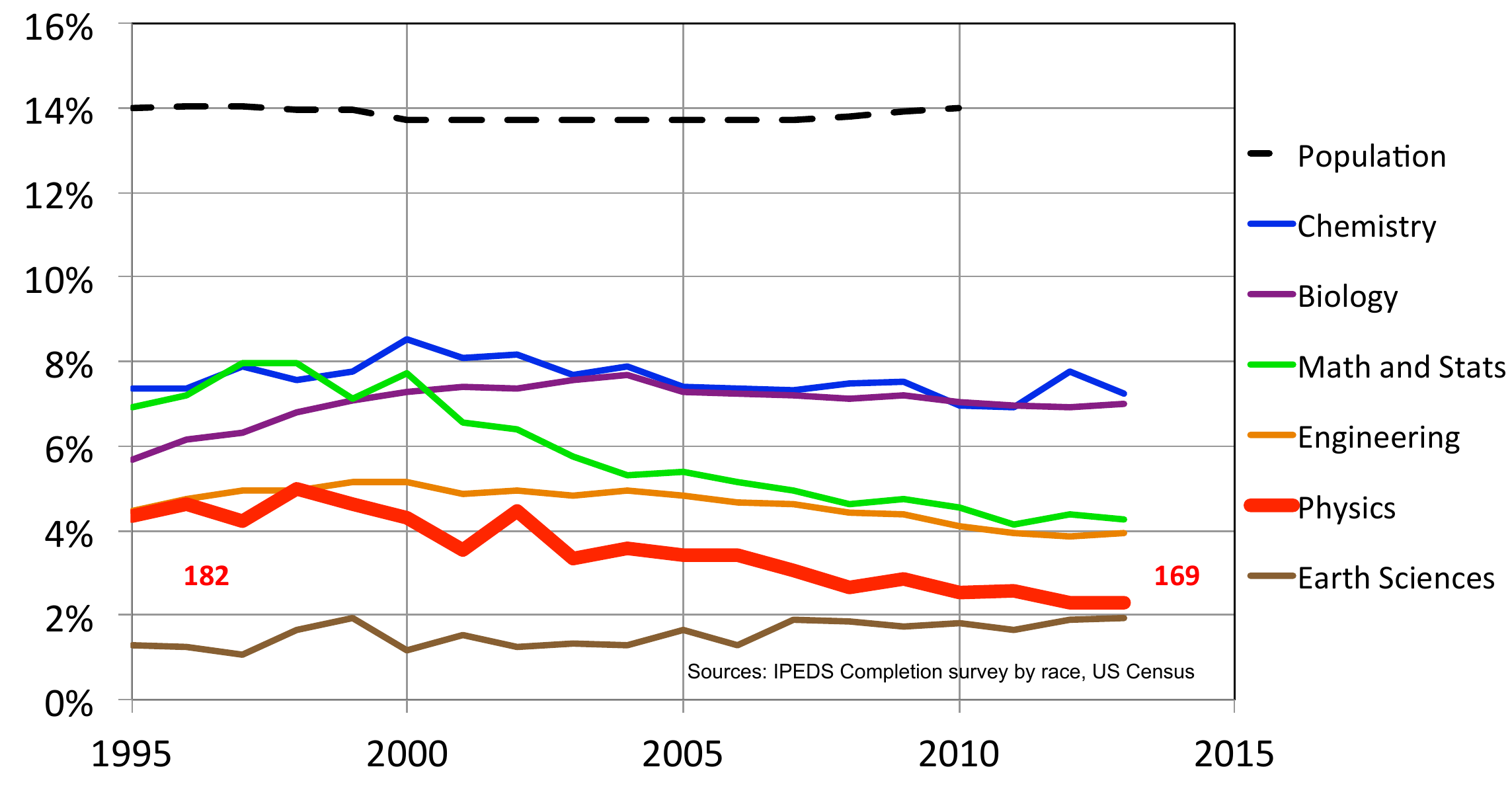}
		\includegraphics*[width=.49\columnwidth]{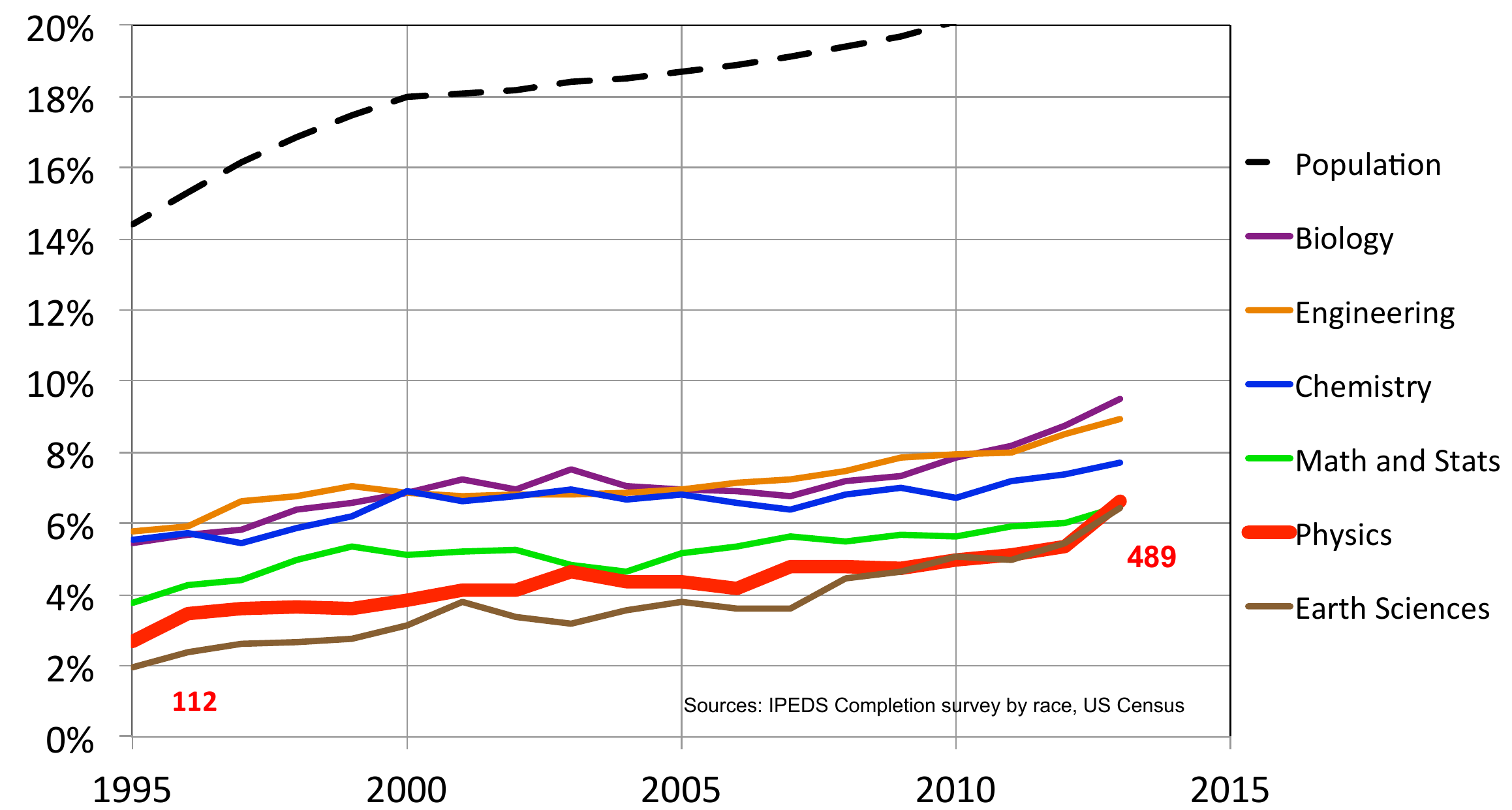}
			\caption{(Color online) Percentage of bachelor's degrees in physical sciences earned by African-American and Hispanic Americans given left to right, respectively~\cite{AIP,APS}.}
  	\vspace{-.2cm}
	\label{fig:urm_physics}
	\end{center}
\end{figure}

As briefly discussed, the percentage of bachelor's degrees earned by URM in physics continues to be close to the bottom for physical sciences. The number earned by Hispanic Americans has shown an increasing trend over the last ten years, but this growth seems to be proportional to the population growth and thus not a clear indicator of increased participation. For African Americans the percentage of bachelor's degrees earned has been steadily falling over the last 10-15 years as shown Figure~\ref{fig:urm_physics}. This has been compounded with the closing of some physics department at Historically Black Colleges and Universities (HBCUs). Hispanic-, African-, and Native -American account for roughly 10$-$11\% and 6\% of BS and PhDs degrees earned in physics~\cite{AIP,APS}. The general trend for bachelors and doctoral degrees earned by URMs over the last 20 years is shown in Figure~\ref{fig:urm_physics}. 

  \begin{figure}[htb]
	\begin{center}
		\includegraphics*[width=.680\columnwidth]{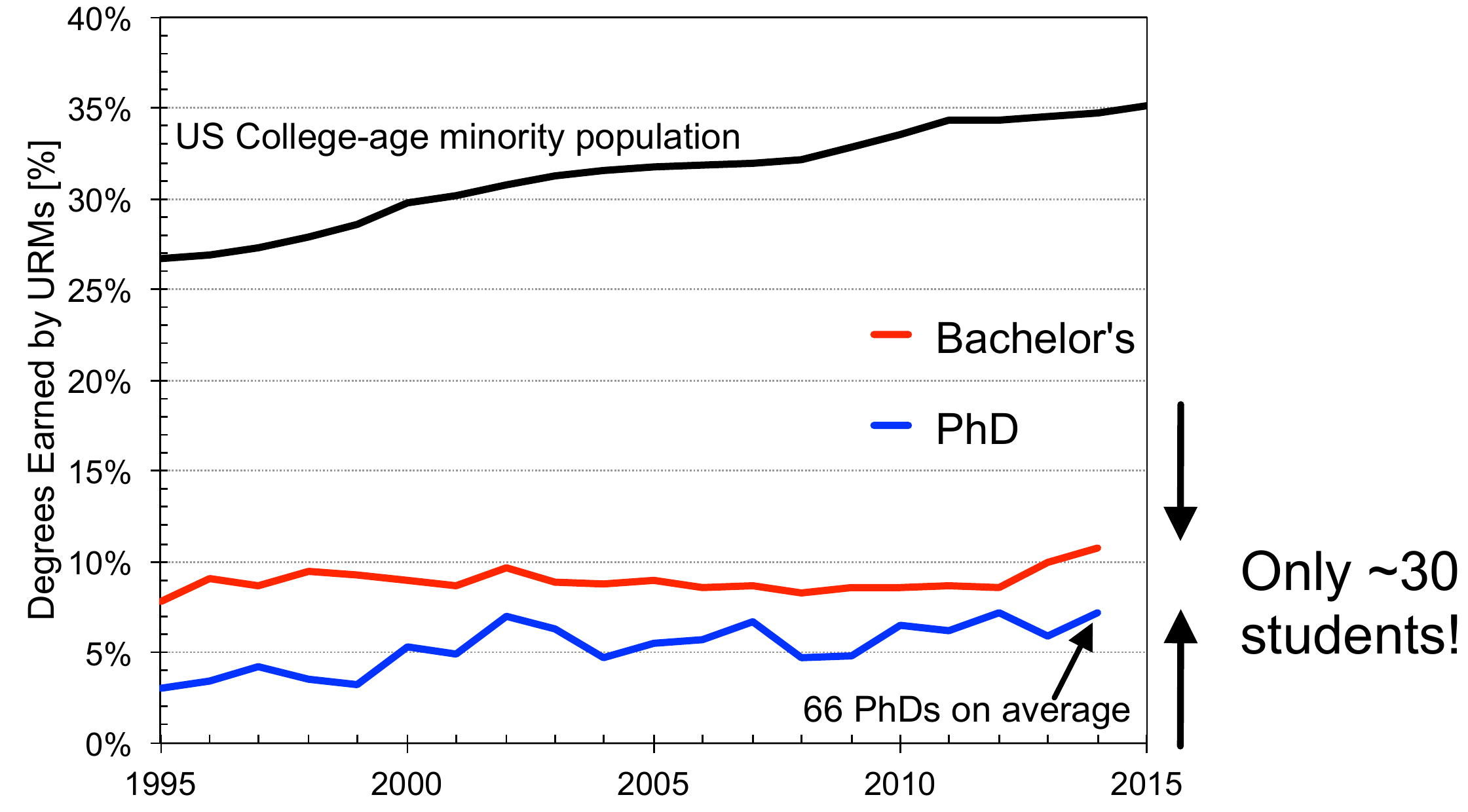}
			\caption{(Color online) Percentage of bachelor's and doctoral degrees earned by URM in physics. The difference between BS and PhD degrees is roughly 30 more earned degrees at the doctoral level. The APS Bridge Program aims to close the gap between the percentage of physics PhDs awarded to underrepresented minority students to match the percentage of physics Bachelor's degrees. }
  	\vspace{-.2cm}
	\label{fig:urm_physics}
	\end{center}
\end{figure}

\section{Bridge Programs}
Bridge Programs are an innovative approach to addressing the underrepresentation of some groups  in physics. They aim to provide opportunities for students to be successful that may not have had such chances by traditional means. The APS Bridge Program has the goal to increase, within a decade, the percentage of physics PhDs awarded to underrepresented minority students to match the percentage of physics Bachelor's degrees granted to these groups as shown in Figure~\ref{fig:urm_physics}. It also aims to develop, evaluate, and document sustainable model bridging experiences that improve the access to and culture of graduate education for all students, with emphasis on those underrepresented in doctoral programs in physics.

\begin{figure}[htb]
	\begin{center}
		\includegraphics*[width=.75\columnwidth]{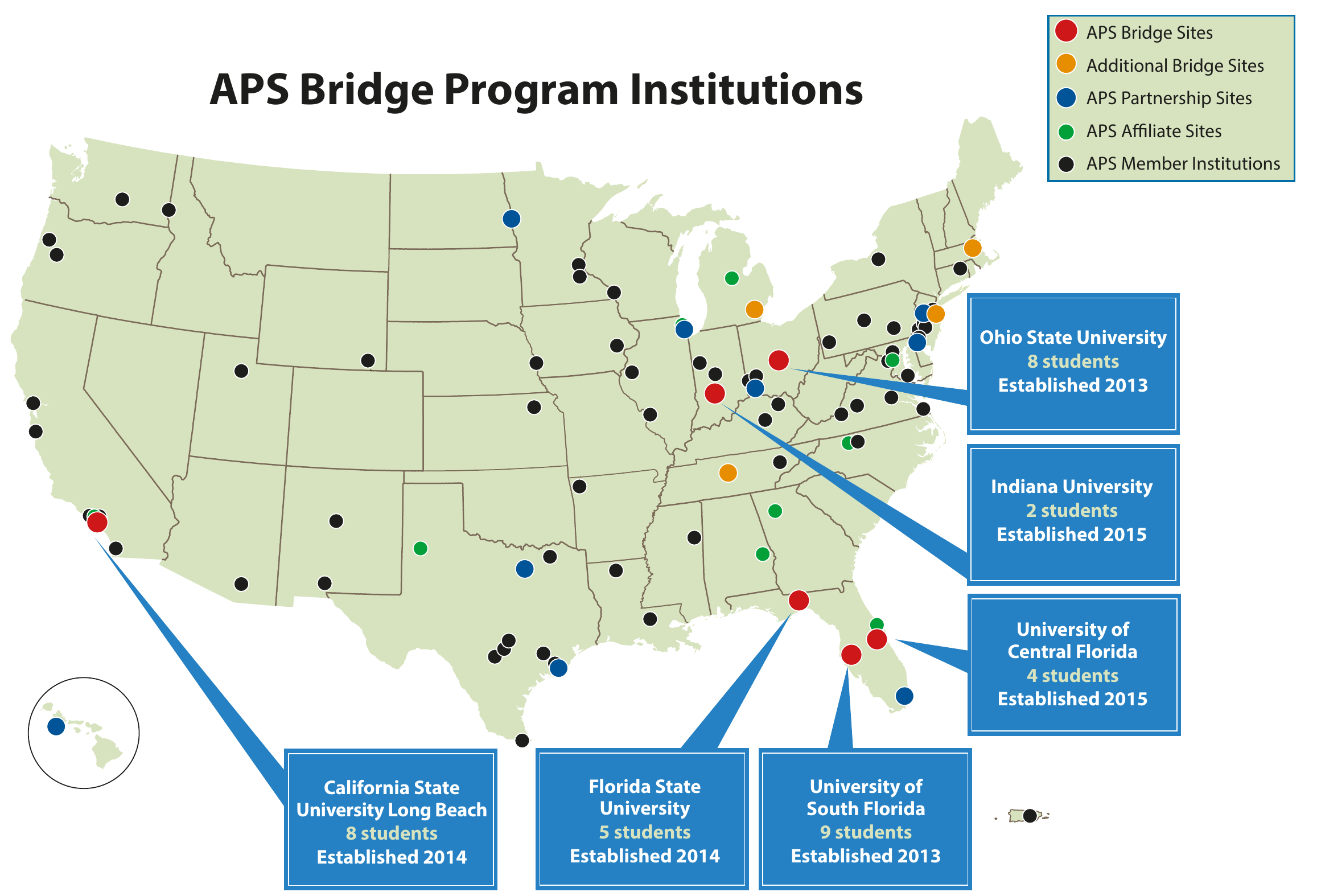}
			\caption{(Color online) National network of institutions participating in and affiliated with the APS Bridge Programs.~\cite{APS_Connections}. The six APS funded bridge sites established in Ohio, Florida, Indiana, and California are indicated by the blue call-out boxes. }
  	\vspace{-.2cm}
	\label{fig:BP_Sites}
	\end{center}
\end{figure}

The project has established six bridge sites in Indiana, Ohio, Florida, and California that provide coursework, research experiences, and substantial mentoring for students who either did not apply to graduate school, or were not admitted through traditional graduate school admissions. These institutions receive funding to develop programs that prepare students to be accepted into a physics doctoral program. Usually, a bridge program accepts students who typically would not gain or did not gain acceptance into a doctoral program for 1-2 years, enhancing their academic and research skills before applying to a doctoral program. There are a few established Bridge Programs that include the Fisk-Vanderbilt bridge program, which was started around 2004, the Imes-Moore Bridge Program in Applied Physics at University of Michigan, Columbia university's bridge program, and the program at MIT. Presently, there are a few more programs under development at institutions such as Princeton and University of Chicago. Physics departments nationwide that participate with the APS Bridge Program can be seen in Figure~\ref{fig:BP_Sites}. The are currently 93 Member Institutions in 36 states and 19 Partnership Institutions in 14 different states. This demonstrates the paradigm shift in physics graduate education.

\subsection{Best Practices} Successful Bridge Programs have certain key components and best practices in common, some of which include: (1)  substantial tenured faculty involvement, (2) adopting a more holistic approach to performing graduate admissions, (3) securing financial support for at least one-year of bridging experience that allows students to resolve major gaps in their undergraduate education. (4) providing multiple mentoring resources, (5) being flexible in coursework, (6) sustained progress monitoring, and (7) looking for a great research match.  

\section{Outcomes}
The APS bridge program outcomes boast 24\% female, and 93\% URM participation of students that have been placed on bridge or graduate programs. Of the 93\% URM participation, the representation of African-, Hispanic-, and Native - American is 24\%, 64\%, and 5\% respectively. The national average of female and URM representation in all U.S. physics graduate departments is 20\% and 6\% respectively. 
A highlight of the APS Bridge Program has been the remarkable retention of the placed students, which is 88\%, a stunning 28\% higher than the 60\% national average in physics graduate programs. Retention is defined as students that enter a PhD program in physics and earn their PhD. The higher retention is indicative of the outlined best practices as other programs have similar results. The Fisk-Vanderbilt bridge program has achieved 80\% retention for the students that enter their program, while the University of Michigan's Imes-Moore bridge program in applied physics has surpassed both programs with a 90\% retention rate. 
Fisk University is now recognized as the highest producer of black students with master's degrees in physics, and Vanderbilt has become a leading producer of doctoral degrees in astronomy, physics and materials science earned by URM students~\cite{FV}. 
\label{sec:outcomes}
  \begin{figure}[htb]
	\begin{center}
		\includegraphics*[width=.50\columnwidth]{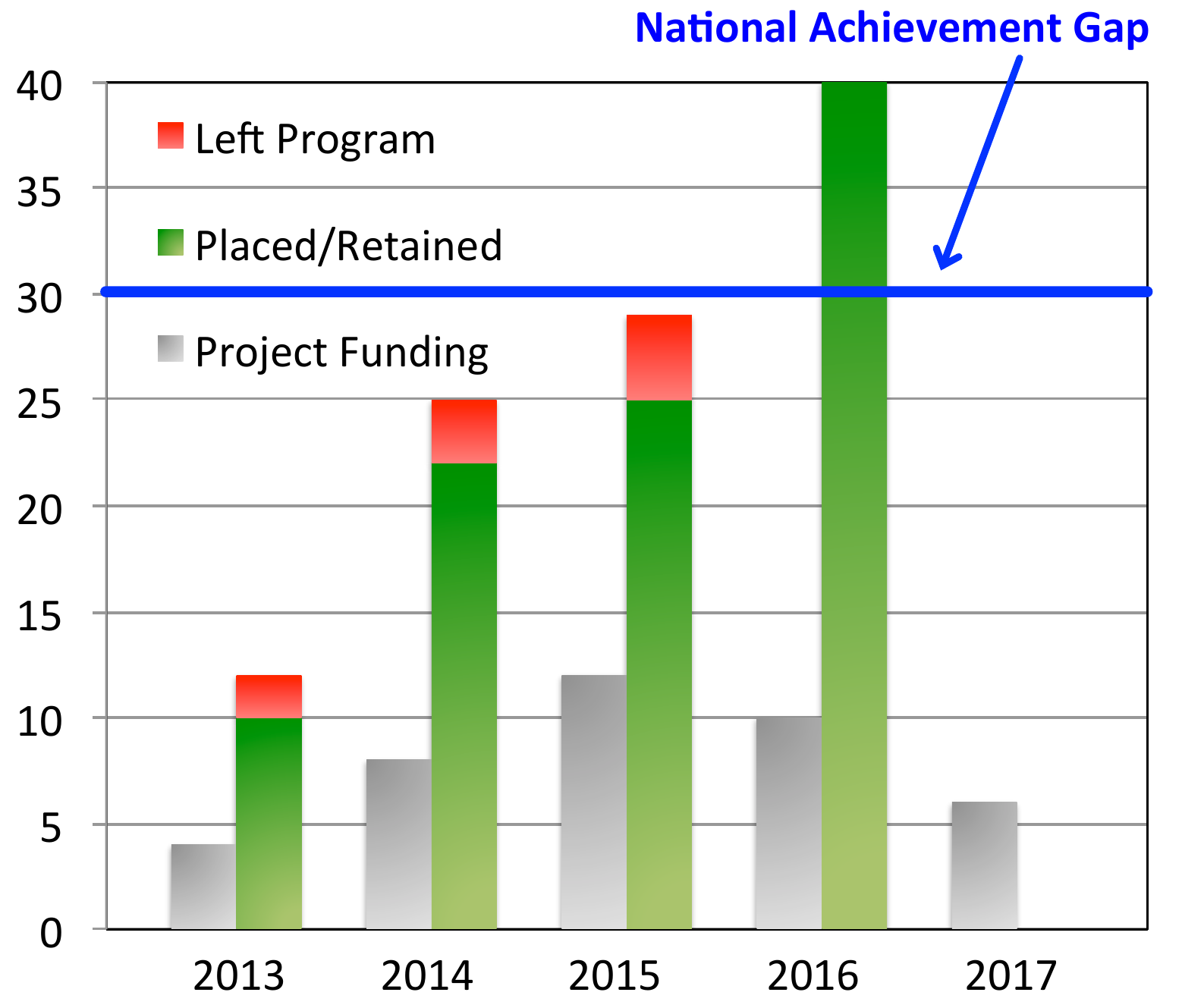}
			\caption{(Color online) Outcomes of APS Bridge Program.}
  	\vspace{-.2cm}
	\label{fig:BP_Outcomes}
	\end{center}
\end{figure}

\section{Conclusions}
The outcomes discussed in section~\ref{sec:outcomes} clearly demonstrate that Bridge Programs can be implemented as an approach to increase diversity in student enrollment, retention, and ultimately the entire physics community. The APS Bridge Program has been influential on placing over 100 students into Bridge or graduate programs in physics while retaining 88\% of those placed. The project achievement since starting is shown in Figure~\ref{fig:BP_Outcomes}. Leaders at physics departments that have Bridge Programs have acknowledged that the bridge programs best practices are beneficial to all student not only those enrolled as bridge students. If we care about science, then we must also care about scientists and Bridge Programs can provide the access and opportunity for students to be successful.

\section{Acknowledgements}
This material is based upon work supported by the National Science Foundation under Grant No. 1143070. Any opinions, findings, and conclusions or recommendations expressed in this material are those of the author(s) and do not necessarily reflect the views of the National Science Foundation.


\begin{thebibliography}{99}
\bibitem{AIP} American Institute of Physics Statistical Research Center (www.aip.org/statistics)
\bibitem{APS} American Physical Society (www.aps.org/programs/education/statistics)
\bibitem{APS_Connections} APS Bridge Program Connections newsletter, Vol. 3,  2016
\bibitem{FV} K. Powell; Higher education: On the lookout for true grit Nature Vol. 504, 2013


\end{thebibliography}
\end{document}